\documentclass[screen,sigplan,nonacm]{acmart}

\usepackage[utf8]{inputenc}
\usepackage{graphicx}
\graphicspath{ {./} {../}}

\usepackage{graphicx}

\usepackage{pgfplots}
\pgfplotsset{compat=1.18}

\usepackage{tabularx}
\usepackage{makecell}
\usepackage{tabu}
\usepackage{enumitem}
\usepackage[export]{adjustbox}
\usepackage{xcolor}
\usepackage{boldline}
\usepackage{textgreek}
\usepackage{multirow}
\usepackage{bigstrut}
\usepackage{xspace}
\usepackage[caption=false]{subfig}
\usepackage{textcomp}
\usepackage{hyperref}
\hypersetup{
    colorlinks=true,
    linkcolor={red!50!black},
    citecolor={green!60!black},
    urlcolor={blue!80!black}}

\usepackage{amsmath,amsfonts}

\usepackage{algorithmicx}
\usepackage{graphicx}
\usepackage{textcomp}
\usepackage{xcolor}

\usepackage{booktabs}
\usepackage[capitalise,nameinlink,noabbrev]{cleveref}
\def\BibTeX{{\rm B\kern-.05em{\sc i\kern-.025em b}\kern-.08em
    T\kern-.1667em\lower.7ex\hbox{E}\kern-.125emX}}

\usepackage{fancyhdr}

\usepackage{tikz}
\newcommand*\circledblue[1]{\tikz[baseline=(char.base)]{
            \node[shape=circle,draw=blue, thick,inner sep=1pt] (char) {\textbf{\textcolor{blue}{#1}}};}}
\newcommand*\circledorange[1]{\tikz[baseline=(char.base)]{
            \node[shape=circle,draw=orange, thick,inner sep=1pt] (char) {\textbf{\textcolor{orange}{#1}}};}}

\newcommand*\circledgreen[1]{\tikz[baseline=(char.base)]{
            \node[shape=circle,draw=darkgreen, thick,inner sep=1pt] (char) {\textbf{\textcolor{darkgreen}{#1}}};}}

\usepackage{xcolor,pifont}
\newcommand*\colourcheck[1]{
  \expandafter\newcommand\csname #1check\endcsname{\textcolor{#1}{\ding{52}}}
}
\colourcheck{blue}
\colourcheck{green}
\colourcheck{red}

\newcommand{\parhead}[1]{\vspace{1pt plus 2pt minus 1pt}\par\noindent\textbf{#1}\hspace{.4em plus .2em minus .2em}}

\newcommand{\colorcomment}[2]{\leavevmode\unskip\relax}

\definecolor{darkviolet}{HTML}{9400D3}
\definecolor{darkgreen}{rgb}{0,0.65,0}

\newcommand{\pcie}[1]{\texttt{#1}}

\newcommand{\myname}[1]{Thunderhammer}

\title[Thunderhammer: Rowhammer Bitflips via PCIe and Thunderbolt (USB-C)]{Thunderhammer: Rowhammer Bitflips via\\ PCIe and Thunderbolt (USB-C)} 

\makeatletter
\newcommand{\linebreakand}{
  \end{@IEEEauthorhalign}
  \hfill\mbox{}\par
  \mbox{}\hfill\begin{@IEEEauthorhalign}
}
\makeatother

\author{Robert Dumitru}
\authornote{Equal contribution first author.}
\affiliation{
  \institution{Ruhr University Bochum \&\\ The University of Adelaide}
   \country{}
  }
\email{robert.dumitru@adelaide.edu.au}

\author{Junpeng Wan}
\authornotemark[1]

\affiliation{
  \institution{Purdue University}
   \country{}
  }
\email{wan155@purdue.edu}

\author{Daniel Genkin}
\affiliation{
  \institution{Georgia Tech}
   \country{}
  }
\email{genkin@gatech.edu}

\author{Rick Kennell}
\affiliation{
  \institution{Purdue University}
   \country{}
  }
\email{rick@purdue.edu}

\author{Dave (Jing) Tian}
\affiliation{
  \institution{Purdue University}
   \country{}
  }
\email{daveti@purdue.edu}

\author{Yuval Yarom}
\affiliation{
  \institution{Ruhr University Bochum}
   \country{}
  }
\email{yuval.yarom@rub.de}

\AtBeginDocument{
  \providecommand\BibTeX{{
    Bib\TeX}}}

\begin{document}

\thispagestyle{empty}

\pagestyle{empty}

\begin{abstract}
In recent years, Rowhammer has attracted significant attention from academia and industry alike.
This technique, first published in 2014, flips bits in memory by repeatedly accessing neighbouring memory locations.
Since its discovery, researchers have developed a substantial body of work exploiting Rowhammer and proposing countermeasures. 
These works demonstrate that Rowhammer can be mounted not only through native code, but also via remote code execution, such as JavaScript in browsers, and over networks. 

In this work, we uncover a previously unexplored Rowhammer vector. 
We present \textbf{Thunderhammer}, an attack that induces DRAM bitflips from malicious peripherals connected via PCIe or Thunderbolt (which tunnels PCIe).
On modern DDR4 systems, we observe that triggering bitflips through PCIe requests requires precisely timed access patterns tailored to the target system. 
We design a custom device to reverse engineer critical architectural parameters that shape PCIe request scheduling, and to execute effective hammering access patterns. 
Leveraging this knowledge, we successfully demonstrate Rowhammer-induced bitflips in DDR4 memory modules via both PCIe slot connections and Thunderbolt ports tunnelling PCIe.

\end{abstract}

\maketitle

\section{Introduction}

Rowhammer is a fault injection attack that exploits a hardware vulnerability in modern DRAM: repeatedly accessing or ``hammering'' rows in system memory can disturb and flip bits in adjacent rows.
The disturbance error mechanism responsible for this phenomenon was first discovered in 2014~\cite{kim2014flipping}.
Soon after in 2015, \citet{seaborn2015exploiting} 
showed it could be weaponised to bypass memory protection, escalate privileges to gain kernel-level permissions, and escape sandboxed environments.
In the decade since, Rowhammer has garnered significant attention 
from both academia and industry.

Several defence mechanisms have been proposed at various layers of hardware and software,
however none of them is a panacea, and virtually all impose some performance overhead. 
Error-Correcting Code (ECC) DRAM could reduce the Rowhammer attack surface as it can detect and correct a limited number of bitflips.
However, attacks can still be reliably mounted as a previous study shows~\cite{cojocar2019exploiting}. 
Beyond ECC, modern DDR4 systems and onward incorporate Target Row Refresh (TRR) within DRAM chips or memory controllers (e.g., Intel’s pseudo‑TRR)~\cite{frigo2020trrespass}.
TRR works by monitoring row activation counts to identify potential aggressor rows, then once a predefined threshold is exceeded, refreshing adjacent rows at the next scheduled refresh cycle. 
Despite these mitigation efforts, with TRRespass, \citet{frigo2020trrespass} demonstrated this protection can be overcome by using multi-sided hammering patterns that overwhelm the activation count samplers. 
Another approach is to increase the DRAM refresh rate, which shortens the vulnerability window for bitflips. 
However, this approach introduces substantial performance and energy overhead.

\begin{figure}[t]
    \centering
    \vspace{0.5em}
    \includegraphics[width=0.47\textwidth]{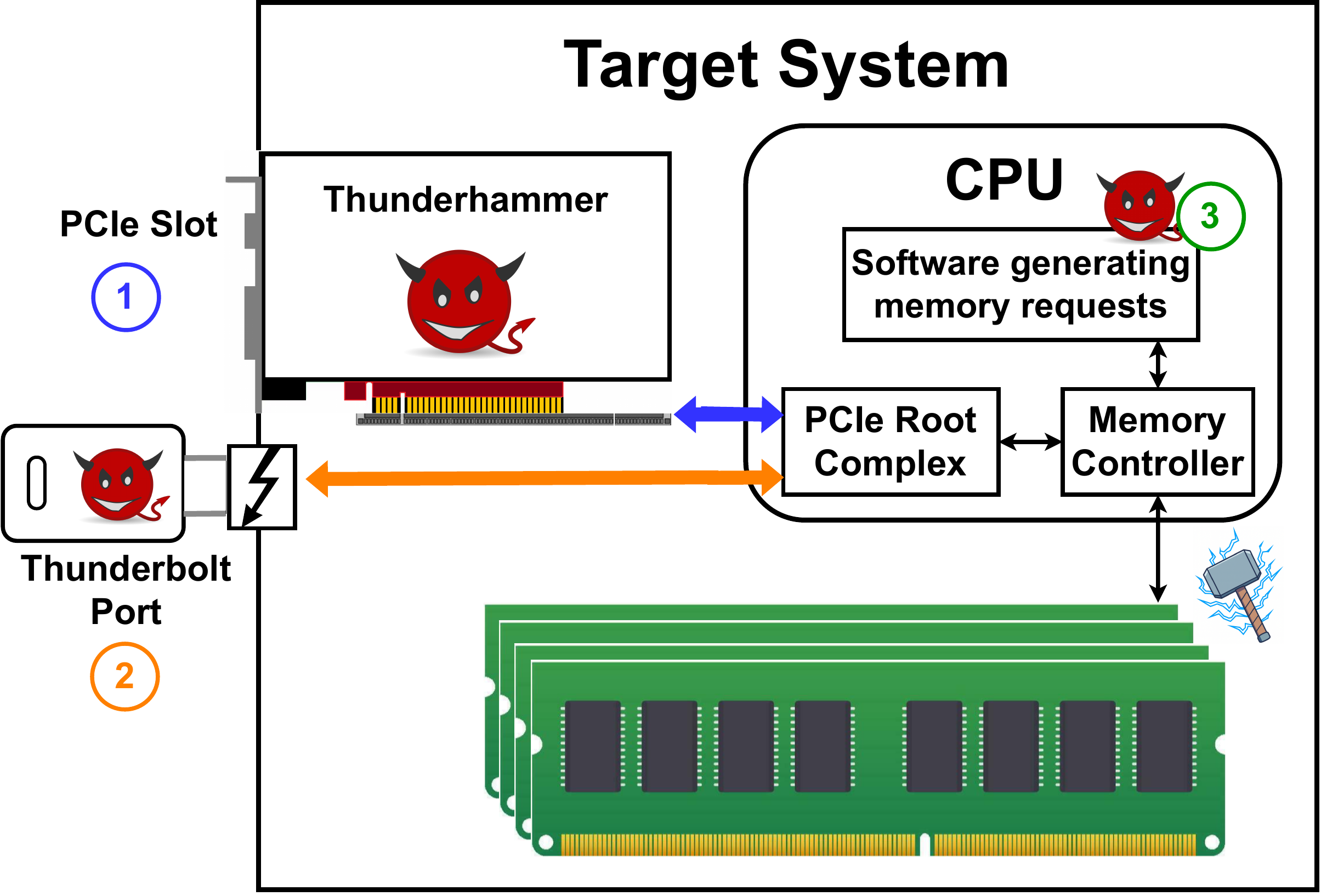}
    \caption{Rowhammer attack vectors: \protect\circledblue{1} Thunderhammer via PCIe slot, \protect\circledorange{2} Thunderhammer via Thunderbolt port tunnelling PCIe, and \protect\circledgreen{3} Typical attack mounted from software.} 
    \label{fig:highlevel}
\end{figure}

Rowhammer attacks have typically been demonstrated from low-privileged code running on the target machine, and in various scenarios such as from virtual machines~\cite{chen2025hyperhammer, xiao2016one} and browsers~\cite{gruss2016rowhammer, de2021smash, frigo2018grand}. 
These make up the typical software-based attacks, illustrated in \protect\circledgreen{3} of ~\cref{fig:highlevel}.

There have also been a few Rowhammer demonstrations from peripherals, namely Nethammer~\cite{lipp2020nethammer} (invoking software-based accesses through NICs), Throwhammer~\cite{tatar2018throwhammer} (exploiting RDMA), and JackHammer~\cite{weissman2019jackhammer} (through heterogeneous CPU-FPGA platforms). 
We presume the first two use network devices connected through PCIe lanes, though neither of the works explicitly mention it. 
All of these works use double-sided hammering, finding that to be sufficient against the targeted systems which mostly use DDR3 system memory. 
However, among these past works, only Nethammer overcomes a form of TRR in DDR4, albeit one that is susceptible to double-sided hammering. 

Given the advent of modern DDR4 systems against which only multi-sided hammering is an effective means of inducing bitflips, in this paper we primarily set out to investigate the feasibility of mounting Rowhammer against these systems using peripherals. 
To our knowledge, no previous work has demonstrated this capability. 
Moreover, we explore whether such hammering is possible via generic PCIe connections, including PCIe tunnelled through external Thunderbolt ports. 
This lends the approach to a variety of device configurations and exploitable scenarios, since hosts generally allocate memory differently depending on the attached device type. 
Thunderbolt compatibility also makes for a considerably relaxed physical access threat model, where an attacker does not need to open the target machine for direct access to the motherboard. 
In this paper we ask the following questions:
\\\\
\noindent \emph{Can Rowhammer be mounted by PCIe peripherals 
against modern DDR4 systems where multi-sided hammering is necessary to overcome TRR? 
Could off-the-shelf devices be used for such attacks?
Can hammering functionality be built into generic PCIe and Thunderbolt peripheral devices?}

\subsection{Our Contribution}
In this paper we show that multi-sided
Rowhammer against modern DDR4 from generic peripherals is possible when using specifically tuned hammering patterns. 
We propose \textbf{\myname{}}, an attack wherein an adversary that only has control over a generic device connected through PCIe or Thunderbolt can cause Rowhammer bitflips, thereby violating security guarantees predicated on memory isolation. 
In principle, this opens the possibility of circumventing memory isolation mechanisms that IOMMU enforce for peripherals. 
We illustrate the \myname{} attack vectors \protect\circledblue{1} / \protect\circledorange{2} in~\cref{fig:highlevel}, distinguishing them from typical software-mounted attacks \protect\circledgreen{3}.

As a result, \myname{} demonstrates the potential for memory isolation breakout 
by malicious PCIe and Thunderbolt peripherals against a wider range of targets.
This makes an already dangerous attack even more powerful, as through Thunderbolt (USB-C connector), Rowhammer inherits the classic malicious-USB threat model.

\parhead{Summary of Contributions.}
In this work, we make the following contributions:
\begin{itemize}[nosep,left=0pt]
    \item We demonstrate Rowhammer against DDR3 targets using simple, off-the-shelf software-controllable PCIe devices both connected via a PCIe slot and a Thunderbolt 2 port  
    \item We build a custom PCIe hammering device capable of issuing memory requests with precise timing, and of monitoring a target's memory controller throttling behaviour
    \item Using our custom device hardware, we investigate the interaction of our hammering request flows with the buffering mechanisms of a DDR4 target memory controller
    \item We successfully demonstrate Rowhammer using multi-sided hammering against DDR4 targets via both a PCIe slot and a Thunderbolt 3 port, enabled by carefully tuned access patterns derived from our analysis
\end{itemize}

\noindent Our custom device hardware design is openly available at: \hbox{\url{https://github.com/0xADE1A1DE/Thunderhammer}}

\section{Background}
\label{section:backg}

\subsection{PCI Express and Thunderbolt}

PCI Express (PCIe) is the standard protocol used to connect peripheral devices to CPUs from manufacturers including Intel, AMD, and ARM. 
PCIe inherits compatibility from the earlier PCI standard. 
Architecturally, PCIe functions as a high-speed serial network, supporting point-to-point connections. 
The PCIe protocol consists of three primary layers: the Physical Layer, the Data Link Layer, and the Transaction Layer. 
In \myname{} attacks, our peripheral devices attack the host by accessing main memory at a high frequency using crafted Transaction Layer Packets (TLPs).

Data transmission within PCIe is credit-based to ensure efficient flow control. 
Additionally, a single upstream node (such as the CPU) can connect to multiple downstream nodes (peripheral devices). 
To facilitate this, PCIe switches or Platform Controller Hubs (PCH) may be integrated onto the motherboard, providing connections to numerous downstream peripherals.

\parhead{Thunderbolt.}
Initiated by Intel and Apple, Thunderbolt is a high-speed interface widely deployed in modern laptops and compact desktops, and since Thunderbolt 3 it has adopted the USB-C connector, which integrates PCIe, DisplayPort, USB, and power delivery into a versatile single-cable solution. 
Beginning with Thunderbolt\,3, it offers up to 40\,Gb/s of bidirectional bandwidth by multiplexing multiple protocols—including four lanes of PCIe\,3.0 (up to ~32\,Gb/s total), USB 3.1 at 10\,Gb/s, and DisplayPort\,1.2 video streams over a single USB-C cable~\cite{thunderbolt3blog}.

\parhead{IOMMU.}
PCIe peripherals communicate with the host via four primary mechanisms: (1) \textit{port‑mapped I/O (PMIO)}, using \texttt{in/out} instructions in x86 to access a separate I/O address space which is a legacy approach unavailable on ARM or RISC‑V (2)  \textit{MMIO (memory‑mapped I/O)}, the predominant method, in which peripheral registers are mapped into the CPU’s physical memory space and accessed through normal load/store instructions~\cite{mmio_pio_wikipedia}; (3)  \textit{DMA (direct memory access)}, configured via MMIO, which enables PCIe peripherals to transfer data directly to or from system memory without CPU intervention, freeing the CPU to perform other tasks while the transfer is ongoing; (4) \textit{Interrupts}, allow PCIe peripherals to asynchronously notify the CPU of events, such as the \pcie{Completion} of a DMA transfer, using either interrupt lines or  Message-Signalled Interrupts (MSI). 

With DMA support, a malicious or compromised PCIe peripheral can potentially access the entire physical memory, even if only certain DMA regions were intended for use. 
An attacker could thus steal sensitive data or inject harmful code into the system. 
To mitigate such DMA attacks, modern platforms employ an \textit{Input–Output Memory Management Unit (IOMMU)}~\cite{wiki:IOMMU}. 
The mechanism of an IOMMU is similar to that of a Memory Management Unit (MMU). 
While an MMU ensures that a process can only access physical memory addresses mapped to its own virtual address space, an IOMMU ensures that a peripheral device can only access host physical addresses mapped explicitly to its device address space. 
IOMMU support is provided by all major CPU vendors: Intel VT-d (Virtualisation Technology for Directed I/O), and ARM System Memory Management Unit (SMMU).

\parhead{Attacks on PCIe and Thunderbolt.}
Invisible Probe~\cite{tan2021invisible} is an PCIe side-channel attack based on traffic contention. 
By timing the variance of IO latency cased by PCIe traffic, the attackers could capture website fingerprint and password keystrokes.

Launched from a Thunderbolt peripheral, Thunderclap~\cite{markettos2019thunderclap} exploits vulnerabilities in IOMMU misconfiguration, potentially leading to secret leakage and arbitrary code execution. 
Thunderspy~\cite{ruytenberg2022lightning} also leverages a Thunderbolt device to attack Thunderbolt Security Levels. 
In contrast, \myname{} focuses on triggering Rowhammer attacks from a Thunderbolt peripheral.

\subsection{DRAM}
Dynamic Random-Access Memory (DRAM) is organised hierarchically into channels, ranks, banks, rows, and columns. A channel contains multiple ranks, and each rank is further divided into multiple banks. Within a bank, data is stored in a two-dimensional array of cells, with each cell holding a single bit. To access data, an entire row must first be activated and loaded into the \texttt{row buffer}, which is private to each bank. Only when a row is present in the row buffer can column-level accesses be performed.
The memory controller orchestrates the scheduling and ordering of DRAM commands. A typical access sequence, when the desired row is not already open, proceeds as follows:
$$
\texttt{PRE} \;\rightarrow\; \texttt{ACT} \;\rightarrow\; \texttt{RD} \;\rightarrow\; \texttt{PRE}
$$
In the above process, a \texttt{PRE} (precharge) command closes any previously open row and prepares the bank. An \texttt{ACT} (activate) command then loads the target row into the row buffer. Once active, a \texttt{RD} (read) command retrieves data from the specified column. A subsequent \texttt{PRE} may be issued to close the row and ready the bank for another access.
For writes, the sequence is similar but uses a \texttt{WR} (write) command instead of \texttt{RD}. If consecutive accesses target the same open row, the controller can bypass the initial \texttt{PRE}/\texttt{ACT} steps, reducing latency and improving throughput. This condition is known as a \texttt{row hit}. Conversely, if a different row in the same bank must be accessed, the controller issues a precharge followed by activation, leading to a \texttt{row conflict}.

\subsection{Rowhammer}\label{sec:usb-comm}
The continued down-scaling of semiconductor technology has made DRAM cells smaller and packed closer together, making them more susceptible to electrical interference from nearby cells, which caused disturbance errors and Rowhammer attacks~\cite{mutlu2019rowhammer}.
Typical DDR3 Rowhammer attacks can be launched by repeatedly activating one or two rows adjacent to a victim row, known as single-sided and double-sided Rowhammer attacks~\cite{kim2014flipping, seaborn2015exploiting}. 

For DDR4 DIMMs, simple single-sided or double-sided Rowhammer attacks are ineffective due to the Target Row Refresh (TRR) mechanism, which monitors aggressor rows and refreshes potential victim rows. 
TRRespass~\cite{frigo2020trrespass} was the first to propose many-sided attacks, which use multiple aggressor rows to exploit the fact that TRR can only track a limited number of them.
Building on this line of research, Blacksmith~\cite{jattke2022blacksmith} exploits non-uniform aggressor row access patterns to uncover additional bitflips. 
SledgeHammer~\cite{kang2024sledgehammer} further accelerates Rowhammer bitflips on DDR4 by leveraging bank-level parallelism.

Beyond Rowhammer attack from the CPU side, there are also Rowhammer attacks based on PCIe peripherals.
JackHammer~\cite{weissman2019jackhammer} take advantage of an FPGA to attack a host computer by launching Rowhammer attacks and covert channel attacks. 
However, in JackHammer, the host–peripheral interaction is handled through a dedicated software stack, specifically, the Open Programmable Acceleration Engine (OPAE), rather than via raw TLP packets.

Nethammer~\cite{lipp2020nethammer} induces Rowhammer bitflips via a Network Interface Card (NIC), relying on the assumption that the host’s network stack uses uncached memory or performs cache flushes (e.g., via \textit{clflush}) when processing incoming packets. 
Targeting specific application, i.e., RDMA‑memcached, Throwhammer~\cite{tatar2018throwhammer} can induce bitflips via RDMA NICs and is capable of bypassing IOMMU memory isolation.
The recent GPUHammer study~\cite{lin2025gpuhammer} demonstrates bitflips on a GPU device, successfully compromising the NVIDIA A6000 GPU using user-level CUDA code. 

In addition, RowPress~\cite{luo2023rowpress} is another variant of memory attacks that shares mechanisms with Rowhammer. Instead of repeatedly activating aggressor rows, RowPress keeps the aggressor row open for an extended period to amplify read disturbance, which can also induce bitflips in DDR4 systems.

\section{Hammering with Simple PCIe Devices}
\label{sec:simplehammer}

We investigate the viability of triggering Rowhammer bitflips using simple approaches with off-the-shelf PCIe and \hbox{Thunderbolt} devices. 
We consider a scenario wherein an attacker only has control over a device that is physically plugged in to a target machine, but no direct control over code executed on the target machine. 
As shown in \cref{fig:highlevel}, the attack device connects to a target machine in one of two ways: \circledblue{1} a PCIe slot on the motherboard, or \circledorange{2} an original Thunderbolt port or a USB \hbox{Type-C} port that supports 
Thunderbolt. 
For \circledorange{2} we note that the system must support tunnelling PCIe through Thunderbolt, which is common. 

Aside from regular PCIe device functions, the attack device performs a malicious, but permitted function of transmitting PCIe requests to memory within its allocated region.
The attack device sends these requests at a high frequency to cause memory accesses in a hammering pattern that triggers Rowhammer bitflips at memory locations outside of its allocated region to manipulate inaccessible data, thus violating security guarantees predicated on memory isolation. 
The functionality responsible for mounting our attack is compatible with those being built into regular devices, alongside their normal function. 

We first attempt to induce bitflips using off-the-shelf controllable PCIe devices, with some basic modifications. 
Our overall aim is to hammer certain aggressor memory locations by issuing a multitude of \pcie{Read} and/or \pcie{Write} PCIe requests to those locations. 
Before presenting our observations, we describe the experimental setup we use. 

\begin{figure}[t]
    \centering
    \includegraphics[width=0.49\textwidth]{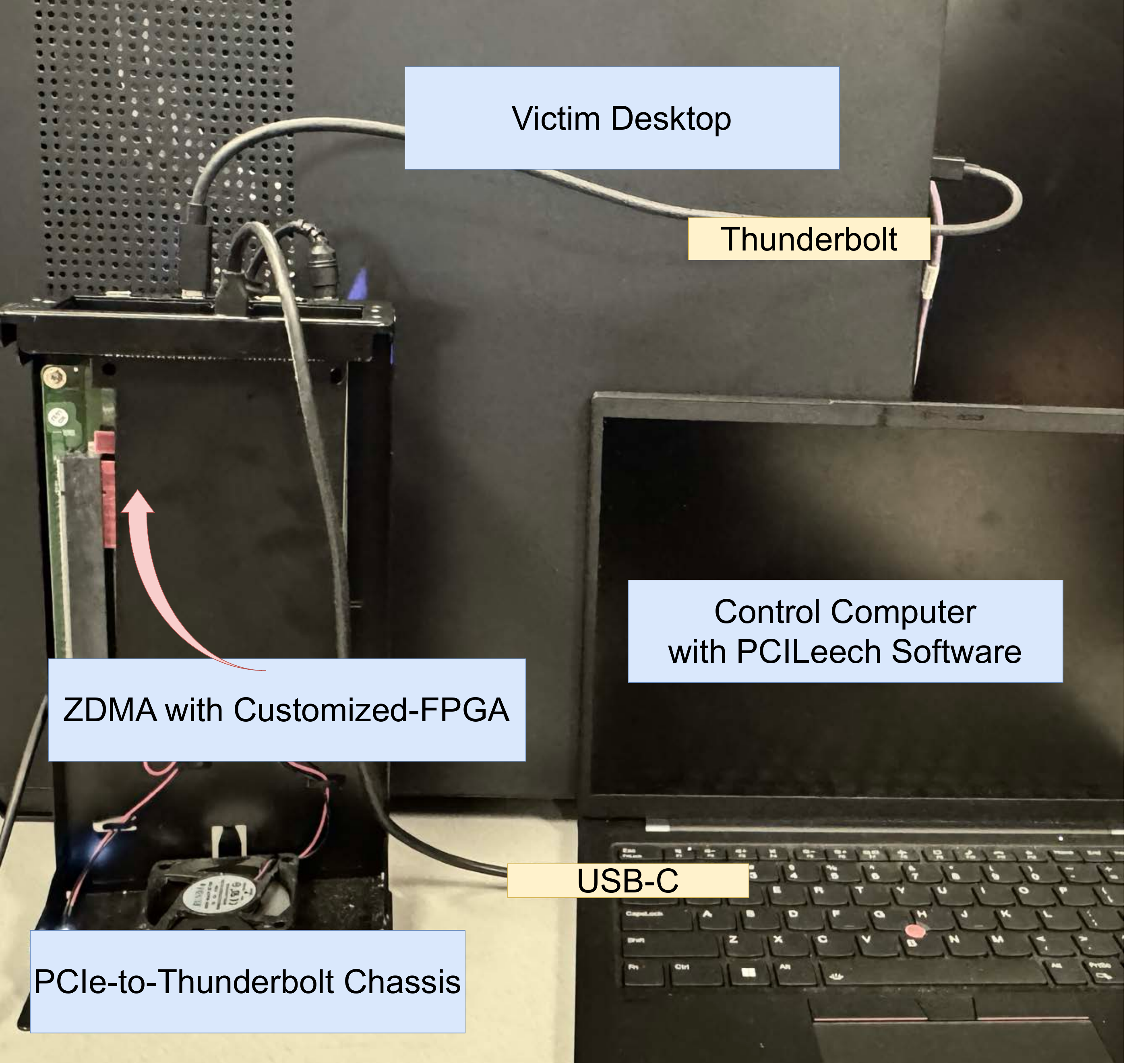}
    \caption{Attack setting for Thunderhammer via a Thunderbolt port tunnelling PCIe}
    \label{fig:settingphoto}
\end{figure}

\subsection{Experimental Setup}
Our hammering device is based on the \emph{pcileech} project~\cite{pcileech}.
This open-source codebase generates commands to dedicated PCIe device hardware from its accompanying \emph{\hbox{pcileech-fpga}} codebase~\cite{pcileechfpgarepo}. 
We use these commands to read and write target system memory using DMA over PCIe. 
Multiple FPGA-based PCIe devices built specifically for use with the \texttt{pcileech} codebase have been produced and made commercially available over time. 
We use the \emph{ZDMA}~\cite{lightningz_zdma}, which is one of the most recent devices and has the highest advertised transfer speeds. 
The ZDMA operates at PCIe Gen\,2 (5\,Gbps lanes) and supports up to x4 lanes. 

\begin{figure}[t]
    \centering
    \includegraphics[width=0.49\textwidth]{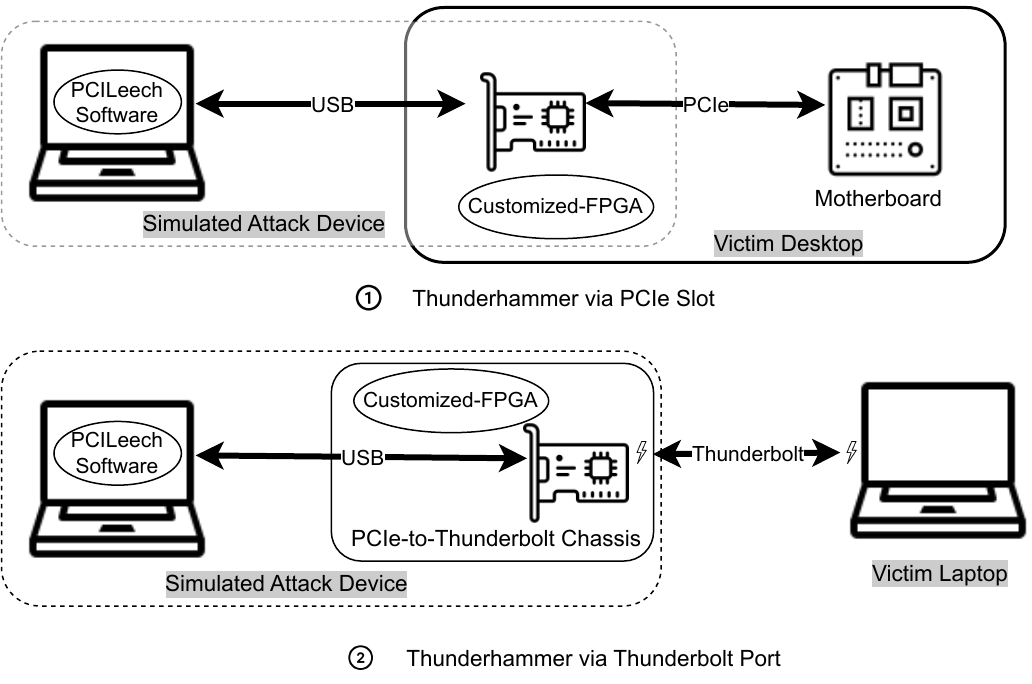}
    \caption{Attack Platform Layouts. \protect\circledblue{1} Thunderhammer via PCIe slot \protect\circledorange{2} Thunderhammer via Thunderbolt port tunnelling PCIe}
    \label{fig:overview}
\end{figure}

The ZDMA, like most other \hbox{pcileech-fpga} devices, is a dual-port device.
On one side is a PCIe connector that connects the board to a slot on target machine, see \cref{fig:overview}.
The other \emph{control} port is a USB 
connection to a machine running the pcileech software. 
As they are provided, 
these devices are designed to only support functionality in target machines with virtualisation and the IOMMU disabled.
In other words, the devices can read and modify all system memory using physical addresses.
For the purposes of testing whether the core mechanism of hammering to trigger bitflips is viable, we
initially 
run our tests against target systems that have virtualisation disabled. 
Under this configuration, our memory requests use the target's physical addressing. 

The pcileech software allows users to issue various types of commands ranging from low-level individually crafted Transport Layer Packets (TLPs) to high-level requests e.g.\ to dump memory from a certain region (in which case the software would then handle crafting all of the necessary TLPs). 
Ultimately, the control software handles these commands and sends just the necessary TLPs to the pcileech device, which it will then transmit over PCIe. 
Among the available user commands is functionality to request looped transmission of certain TLPs.

\parhead{PCIe-to-Thunderbolt.}
To induce bitflips via Thunderbolt, we connect the ZDMA through a PCIe-to-Thunderbolt expansion chassis. 
Such devices are widely available and typically used to connect PCIe devices to computers that have Thunderbolt ports but no exposed or accessible PCIe slots on their motherboard, such as laptops.

\subsection{Bitflips in DDR3 via PCIe and Thunderbolt}
We first experiment with a DDR3 target machine setup.
For an attack from PCIe, we use a Lenovo Thinkcenter desktop with an Intel Core Intel Core i7-4790 CPU, and 4\,GB Samsung M378B5273DH0-CH9 DDR3 system memory, i.e., Desktop2 in ~\cref{tab:platforms}, to which we connect the ZDMA board.
We list all of our target platforms in~\cref{platforms}.
We use a simple double-sided Rowhammer attack from software~\cite{google_rowhammer_test}, verifying that the setup exhibits Rowhammer-induced bitflips.
We then repeat the attack from PCIe, using the ZDMA board, and use software to check whether the content of memory has changed. 
We confirm the presence of bitflips, indicating that attack from PCIe is successful. 

For the Thunderbolt setup, we use an Apple Mac mini i.e., Desktop1 in ~\cref{tab:platforms}, with an Intel Core i7-3615QM CPU, and 4\,GB Samsung M471B5273DH0-CH9 DDR3 system memory.
This system supports Thunderbolt up to version~2 using a Mini DisplayPort connector.
We connect the ZDMA board to the built-in Thunderbolt port, using a Sonnet Echo Express SE II PCIe-to-Thunderbolt expansion chassis~\cite{sonnetecheoii}.
We repeat the attack steps, confirming first that the system is vulnerable to software-induced attacks and then that we can cause bitflips from Thunderbolt.
On our target machine, tunnelling PCIe through Thunderbolt 2 in this scenario works irrespective of whether virtualisation is switched on or off.

We find that using our simple approach we can transmit requests to hammer approximately every 117\,ns, putting it in the same range of performance as Throwhammer~\cite{tatar2018throwhammer} and Nethammer~\cite{lipp2020nethammer} which list their respective access rates as every $\sim$114\,ns and $\sim$163\,ns.

\subsection{Attempting Bitflips in DDR4}\label{section:DDR4attempt}
Next, we attempt to mount Rowhammer via PCIe and Thunderbolt on a DDR4 platform. 
The machine we target is a Gigabyte Z170X-Gaming 7 machine with an Intel Core i7-7700 CPU, and 8\,GB Samsung M378A1K43BB1-CPB DDR4 system memory; listed as Desktop3 in ~\cref{tab:platforms}. 
We use the current state-of-the-art Rowhammer fuzzing technique, Multibank (SledgeHammer)~\cite{kang2024sledgehammer} running natively, to profile our target system through which we find many ground-truth flippy memory locations and corresponding aggressor patterns. 
These patterns include the single-bank case, equivalent to TRRespass~\cite{frigo2020trrespass} fuzzing. 
The targeted memory module implements TRR, and we find that multi-row aggressor patterns that access at least 8 rows are needed to trigger bitflips. 
When simply using the default pcileech software and hardware, unlike with DDR3, we are unable to trigger DDR4 bitflips. 

While pcileech devices like the ZDMA may perform well in terms of throughput, their system is not geared for header-heavy flows of many successive PCIe requests with little to no payload. 
Rather, the system throughput is maximised when transferring bulk data in or out with fewer TLP requests of large payload sizes. 
Since pcileech's looped TLP function is performed in the control software, ultimately all of the individual TLPs containing each memory access command must be sent over USB to the ZDMA for PCIe transmission. 
This subjects them to various stages of buffering as they propagate through the USB connection, USB controller, and other hardware elements before being transmitted over the PCIe lanes, limiting the overall throughput. 
As previously mentioned, we observe transfer rates that let us send 
successive individual requests approximately every $\sim$117\,ns.

\parhead{Looping Transmissions Directly in Hardware.}
To address buffering and bottleneck issues with our approach, we modify the \hbox{pcileech-fpga} hardware core configured onto our ZDMA. 
Specifically, we modify it
to loop the intended TLP transmissions directly in hardware, to the full capability of the PCIe transceiver built into the ZDMA FPGA. 
In detail, we configure the ZDMA to store a sequence of TLPs it receives from the control software, which we call a \emph{batch}, and once it has received the entire sequence it proceeds to repeatedly transmit the batch in a loop, with the packets within a batch sent out immediately one after another.

These modifications allow us to maximise the throughput of header-heavy TLP sequences that are ideal for hammering. 
For the 5\,Gbps Gen\,2 transmission rate that the ZDMA is capable of over x4 lanes, resulting in a speedup of at maximum just over $\times\,6.5$, that amounts to a \pcie{64-bit-address} \pcie{Read} or \pcie{Write} TLP every $\sim$17\,ns. 
This corresponds to a maximum of just over $3.7 \times 10^6$ requests per 64\,ms DRAM refresh period. 
In the case of a one-to-one relation of TLP requests to row activations, it would allow us to do hammering with minimum 50\,K accesses per aggressor for over 50 aggressors, assuming we could sustain transmission at this rate. 

With these modifications we are still not able to trigger DDR4 bitflips. 
Our overall transmission rate should amount to sufficient row activations to induce bitflips, provided enough of the TLP requests issued result in unique row activations.
Nevertheless, we observe no bitflips, which means that something must be limiting our hammering approach.
During the brief periods of max-rate transmission, a large number of requests arrive in a short period of time.
This heightens the opportunities for any request reordering at buffering stages, such as reordering performed by memory controllers to minimise row activations.

\section{Memory Controller Reverse Engineering and Hammering DDR4 Targets}
\label{sec:mcre}

Here, we address the issues with hammering against a DDR4 target system identified in the previous section. 
We turn our attention to investigating how we can avoid reordering effects to maximise the number of row activations caused by our TLPs for hammering. 
We consider the influence of each major element along the chain of connection from our attacker-controlled PCIe device through to the DRAM row buffer. 
As shown in \cref{fig:highlevel}, the major components between the PCIe slot and the target DRAM modules are the PCIe Root Complex and the Memory Controller. 
To mitigate contention-related interference on the Root Complex in our investigation, we perform all of our experiments with no peripherals connected to the other exposed PCIe slots on the target motherboards, and no devices connected to other Thunderbolt ports.
We note that PCIe as a transport layer can reorder requests in-flight.
However, this is only done according to a well-defined set of rules which we can overcome. 
This leaves us to primarily investigate the Memory Controller and its interaction with the target DRAM modules. 

To better understand how reordering and buffering limit hammering, we design experiments to reverse engineer certain functionality and hardware structures within memory controllers. 
In these experiments, we use our controllable PCIe device (ZDMA) as a means of influencing and observing aspects of the memory controller's state. 
This motivates further modifications to our ZDMA device hardware to incorporate the capability of signalling certain events, and high-precision control over the timing of TLP transmissions. 
From our experiments, we find that buffering in certain structures at the target memory controller significantly impacts our ability to induce bitflips in DDR4. 
Using this knowledge and our modified ZDMA hardware\footnote{The source code for our design artefacts is available at \hbox{\url{https://github.com/0xADE1A1DE/Thunderhammer}}}, we are ultimately able to demonstrate bitflips in DDR4 targets from both PCIe and Thunderbolt, using carefully chosen aggressors whose request transmission timing patterns we tune precisely. 
We summarise the optimal timing pattern parameters at the end of \cref{section:DDR4flips}.
In our investigation, we target Intel client-grade CPUs, which affords us both the PCIe and Thunderbolt attack vectors.

\subsection{PCIe Reordering}\label{subsection:pciereorder}

By default, PCIe maintains a strong ordering model across Posted (e.g. \pcie{Write}s) and non-Posted (e.g. \pcie{Read}s) requests~\cite{intel_receive_buffer_reordering}.
Setting a ``Relaxed Ordering'' attribute in TLPs allows requests to bypass each other, therefore to avoid this we ensure the bit is not set in all of our transmitted TLPs.

Conversely, PCIe does not guarantee in-order delivery of \pcie{Completion} TLPs containing data payloads in response to Read requests. 
The PCIe Root Complex may perform reordering after it has been passed the data from DRAM. 
This reordering is generally for the purpose of prioritising smaller data packets, so they are not held up by large-payload \pcie{Completion}s.
We avoid this as much as possible by crafting our \pcie{Read} requests with the minimal possible size.
However, we cannot entirely rely on observing the order of returning \pcie{Completion} to infer the order in which the memory controller issues requests to DRAM.

\subsection{Memory Controller Buffering Mechanisms}

To facilitate the Thunderhammer attack against targets with DDR4 system DRAM, 
we investigate aspects of CPU memory controllers that influence how memory requests arriving from PCIe intended for hammering affect the flow of commands dispatched to DRAM. 
Our analysis focuses on the memory controller architecture in Intel Core CPUs.
We initially consult publicly available documentation, then we conduct targeted experiments to infer the behaviour of undocumented features that affect our attack. 

\begin{figure}[t]
    \centering
    \includegraphics[width=0.49\textwidth]{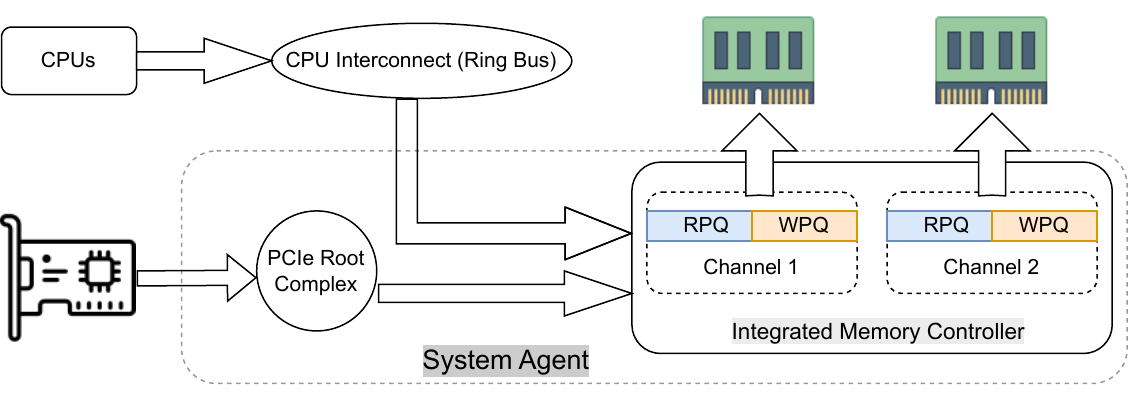}
    \caption{Buffering mechanisms}
    \label{fig:queues-ingress}
\end{figure}

\cref{fig:queues-ingress} shows an overview of our system model. 

\parhead{RPQ and WPQ.}
Memory controllers generally contain separate queues for incoming read requests and write requests. 
According to the Intel Xeon Uncore Performance Monitoring Manual~\cite{intel_uncore_perfmon_2017}, each memory controller includes a Read Pending Queue (RPQ) and a Write Pending Queue (WPQ), which is per channel. 
Although this manual only describes server-grade CPUs, we expect that similarly defined elements and mechanisms are also implemented in their client-grade processors.
To the best of our knowledge, this is not otherwise publicly documented for those systems.

In Intel Core CPUs, the typical interconnect is a ring bus, which connects all CPU cores to the System Agent. 
The System Agent includes the display controller, PCIe interface, and memory controller~\cite{wikichipSkylakeClient}. 
The memory controller is integrated directly into the CPU die and is therefore commonly referred to as the Integrated Memory Controller (iMC)~\cite{wikiMemoryController}. 

The RPQ and WPQ serve as buffering structures: when the memory controller receives Read or Write commands from the cores or PCIe interface,
it allocates entries from the RPQ or WPQ accordingly and later schedules the commands to DRAM.
These queues are credit-based; read or write commands cannot be issued upstream (i.e., from the cores or PCIe) unless there is space available in the corresponding queue.
Once a slot becomes available (i.e., when a command is issued to DRAM and the queue entry is deallocated), new commands can proceed. 

\parhead{Major Modes and Reordering.}
According to~\cite{intel_uncore_perfmon_2017}, the default operating mode of the iMC is \textit{Read Major Mode}.
Reads are typically prioritised over Writes because they are generally more critical for ensuring forward progress.
In contrast, Write requests follow a \textit{posted} approach, meaning a Write is considered complete as soon as it is allocated to the WPQ, without waiting for the data to physically reach DRAM. 

\textit{Write Major Mode} is activated when the occupancy of the WPQ reaches a high watermark.
In this mode, Write requests are prioritised over Read requests.
Additionally, there are two other modes, \textit{Partial Major Mode} and \textit{Isoch Major Mode}~\cite{intel_uncore_perfmon_2017}, which fall outside the scope of our investigation.

Beyond reordering between Reads and Writes, reordering within each group is also possible.
Utilising the WPQ, the iMC allows both Reads and Writes to bypass other Writes that target different addresses.
Before issuing a Read or a Write, the iMC first checks the WPQ to determine if there is an existing Write pending for that specific address.
If a Read request matches an entry in the WPQ, the data can be directly retrieved from the queue instead of accessing main memory.
Similarly, if a Write request matches an existing entry, it will overwrite that data directly~\cite{intel_uncore_perfmon_2017, andreozzi2022milp}.
For each queue, the memory controller is likely to track several requests using a look-ahead window that facilitates reordering
~\cite{intel2023_emif_handbook_v17}.

By understanding the RPQ/WPQ and major mode mechanism, we expect \textit{Read} operations to be more effective at triggering Rowhammer bitflips for the following reasons: 
First, Read requests are prioritised by default in the Read Major mode.
This is corroborated by the description of \texttt{RPQ\_CYCLES\_FULL}, which notes that the RPQ is not expected to become full, except possibly during Write Major mode or very slow DRAM~\cite{intel_uncore_perfmon_2017}.
Thus, using Read requests allows for far more precise control over row activation latency within memory banks than Write requests.
Second, when partial Writes occur (i.e., the data being written to DRAM does not span an entire 64-byte unit which is the default granularity of Writes to DDR4), the memory controller issues additional underfill Reads to retrieve the existing data to perform a read-modify-write~\cite{intel_uncore_perfmon_2017}.
This behaviour complicates the Rowhammer process, whose goal is to trigger highly efficient row activations. 
Although we can issue full-cacheline Write requests from ZDMA to avoid partial Writes, the use of larger TLP packets limits our request transmission rate due to the bandwidth constraints of PCIe.
Finally, PCIe \pcie{Read} requests are non-\textit{posted} and always force a \pcie{Completion} response from the memory system.
This minimises the payload size on \pcie{Read}s transmitted for hammering, so they will have minimal impact on bandwidth.

\begin{center}
    \fcolorbox{black}{gray!10}{\parbox{0.97\linewidth}{
    \textbf{Takeaway 1:}
    Read requests of minimal payloads from PCIe are more effective at triggering Rowhammer bitflips.
    }}
\end{center}

\subsection{RPQ and WPQ Reordering Properties} \label{subsection:reorderexps}
The extent to which the memory controller can reorder a stream of incoming requests depends on five factors:
\begin{itemize}[nosep,left=0pt]
\item the size of its reorder look-ahead window; 
\item how long requests can remain in the look-ahead window without being selected (starvation avoidance);
\item the relative arrival and service rates;
\item the row-level locality; and 
\item the distribution of accesses across banks.
\end{itemize} 
Among these, arrival/service rates, row locality, and bank distribution are determined by the access pattern.
Conversely, 
look-ahead window size and starvation avoidance time are intrinsic properties of the memory controller's internal structure. 
We aim to infer these latter properties through controlled experiments with specifically crafted access patterns. 
Given Takeaway 1, we focus on RPQ hereon.

\begin{figure}[h]
    \centering
    \includegraphics[width=\linewidth]{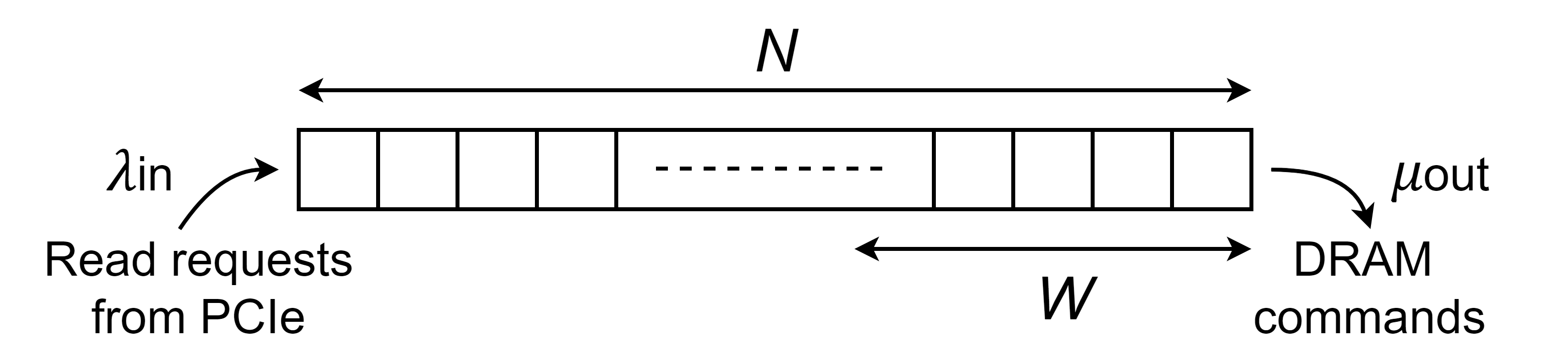}
    \caption{Memory Controller RPQ flows.} 
    \label{fig:RPQ}
\end{figure}

We model an individual queue 
as a bounded queue of depth $N$, with an incoming request arrival rate $\lambda_{\text{in}}$ and a service rate $\mu_{\text{out}}$, see \cref{fig:RPQ}.
The reorder logic acts across a look-ahead window of size $W \leq N$, scanning up to $W$ oldest requests to select the next to be served, delaying some to group same-row requests while still respecting starvation-avoidance limits. 
If $\lambda_{\text{in}}>\mu_{\text{out}}$ over a sufficient period of time, the queue will fill.
When the queue is full, the system applies back-pressure that reaches the PCIe root complex, causing it to throttle peripheral connections. 
We use observations of when this throttling occurs for given combinations of $\lambda_{\text{in}}$, $\mu_{\text{out}}$, and access patterns for some of our reverse engineering experiments. 

\parhead{Modifications to Indicate Throttling.}
To observe throttling as an indicator of an exhausted queue, we modify the ZDMA hardware to signal back to our control software when it has been throttled during packet transmission.

With this mechanism, we find that when we loop PCIe transmissions in hardware, as per our modifications from \cref{section:DDR4attempt}, our logic is regularly interrupted by throttling, after only a few TLP transmissions each time. 
This is because the (in-chip) interface to our Xilinx PCIe transceiver core (PHY)~\cite{PHYUG} is 128 bits wide and clocked at 125\,MHz, meaning we can send a \pcie{64-bit-address Read} TLP (128 bits in length) every 8\,ns. 
This is faster than the speed that our x4 5\,Gbps PCIe Tx lanes support, when factoring in PCIe's 8b/10b encoding and lower link-layer packet overhead. 
This also explains why we arrive at the packet transmission rates listed in \cref{section:DDR4attempt} (one \pcie{64-bit-address Read} per $\sim$17\,ns). 

\begin{figure}[h]
    \centering
    \includegraphics[width=\linewidth]{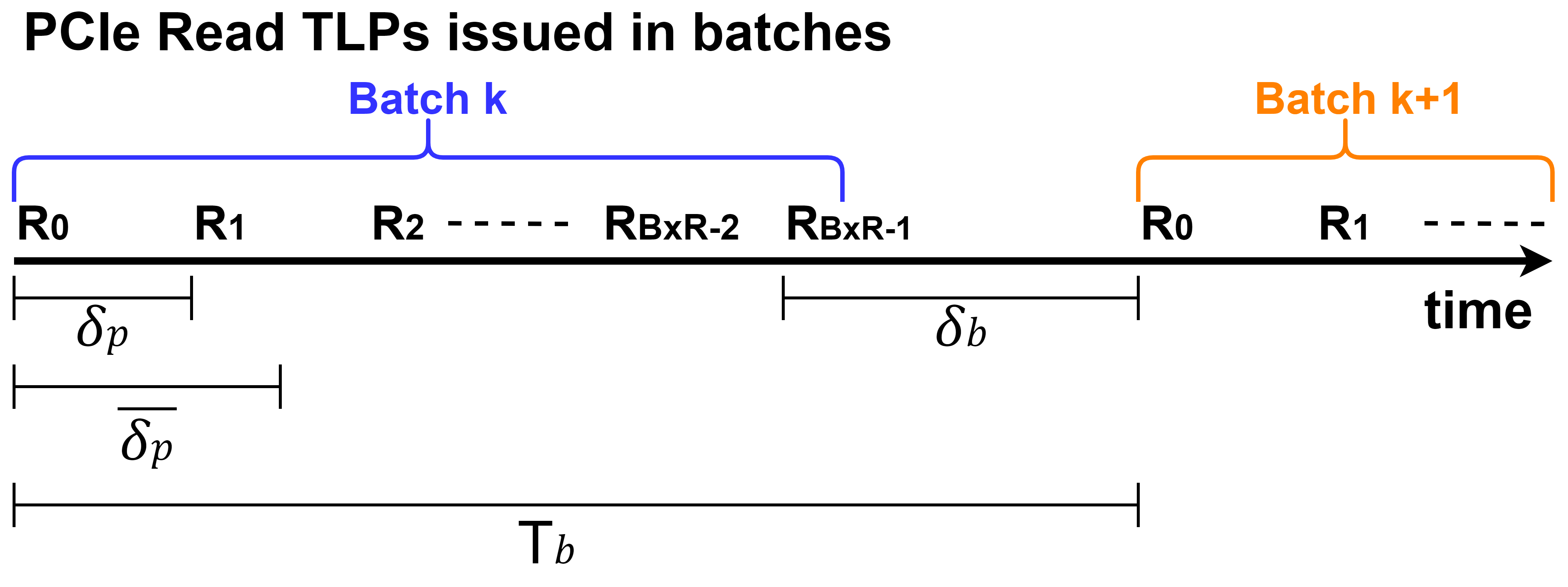}
    \caption{Packet sequence timing parameters for stream of repeatedly transmitted batches of PCIe \pcie{Read} requests.} 
    \label{fig:timing}
\end{figure}

\parhead{Modifications for Precise Transmission Intervals.}
We further modify the ZDMA hardware to transmit TLPs in a loop, with well-defined delay intervals between transmissions that we can set per our core's clock period fidelity of 8\,ns. 
We denote this \emph{inter-packet delay} within a batch as $\delta_{p}$. 
\cref{fig:timing} shows this and the following parameters we describe. 
We also introduce a separate configurable \emph{inter-batch delay}, $\delta_{b}$, between transmission of the last request in a batch and the first in the next batch.
This can similarly only be set according to 8\,ns increments.
We label the batch period $T_b$. 
Now we can transmit batches of requests 
in a loop with precise control over the overall average inter-packet delay of a stream across many batches, i.e., the direct inverse of the incoming request arrival rate, which we denote as $\overline{\delta_{p}} = 1/\lambda_{\text{in}}$.

\parhead{Calibration to Maximum Stable Transmission Rate.}
To use our custom ZDMA's throttling signal as a reliable indicator of exhausted queues, we need to rule out the false positives that arise from our Tx lanes being overwhelmed. 
We first find the maximum transmission rate that the lanes can sustain indefinitely, i.e., the absolute maximum $\lambda_{\text{in}}$ we can use reliably. 
For that, we repeatedly transmit a single \pcie{Read} TLP to the same address in a loop for a long period of time. 
We use a single \pcie{Read} TLP to avoid exhausting the queue. 
We expect repeated requests to the same location to maximise $\mu_{\text{out}}$, since the controller can service many of the Reads following a single row activation. 
In such an idealised scenario, this access pattern would result in the memory controller issuing an \texttt{ACT} command only once for the first request.
To serve every subsequent request, the controller issues only one \texttt{RD} command for each. 
According to the DDR4 JEDEC Standard~\cite{jedec_ddr4_2020}, these can be issued successively every 4\,ns for our type of target DIMM, therefore we have a bound of $\mu_{\text{out}}<1/(4\,\text{ns})$.
We find that transmitting a Read TLP every 16.2\,ns on average \hbox{($\overline{\delta_{p}} = 16.2\,\text{ns}$)} gives the maximum stable $\lambda_{\text{in}}$.
At this rate there is no throttling related to the PHY and the capacity of its lanes. 
The maximum rate we find is well below our expected service rate upper bound, $\mu_{\text{out}}>\lambda_{\text{in}}$, therefore, we are confident the throttling we observe at this limit is due to lane capacity and not the queue filling.

\parhead{RPQ Reordering Look-Ahead Window.}
We design an experiment to estimate the size of RPQ's reordering look-ahead window ($W$).
As mentioned in \cref{subsection:pciereorder}, we know that when we issue \pcie{Read} TLPs from our device, the  \pcie{Completion} TLPs arriving at our device in response can be reordered by more elements on the chain of connection than just the memory controller. 
Since we cannot entirely rely on this for direct observation of reordering we must combine our analysis with an alternative method. 
When reordering occurs, the system can achieve a high $\mu_\text{out}$, i.e. it can serve requests at a fast rate. 
Whereas when there is no reordering, the system will exhibit worst-case performance, i.e., going through the entire cycle of opening and closing DRAM rows for each individual request. 
Following from this, the core idea of our approach is to stream in requests and observe the maximum serviceable request rate ($\mu_{\text{out}}$).  
We then use that as a direct proxy for how much reordering the system is doing on our stream of requests.

\begin{figure}[h]
\centering
\begin{tikzpicture}
\begin{axis}[
  scale only axis,
  width=0.35\textwidth,
  height=0.24\textwidth, 
  xmin=0, xmax=32,
  ymin=8, ymax=64,
  xlabel={No. of requests to unique rows in batch ($R$)},
  ylabel={Min. avg. inter-packet delay $\overline{\delta_{p}}$},
  ytick distance=8,
  xtick distance=4,
  ymajorgrids,
  xmajorgrids,
  yticklabel=\pgfmathprintnumber{\tick}\,ns
]
\addplot+[
  mark=*,
  thick
] coordinates {
  
  (1,16.2) (2,16.2) (3,16.2) (4,16.2) (5,16.2)
  (6,16.2) (7,16.2) (8,18) (9,18) 
  (10,19.2)
  (11,19.64) (12,19.33) (13,20.31) (14,20.6) (15,23) (16,24)
  
  (17,25) (18,26) (19,27) (20,28) (21,32)
  
  (22,50) (23,50) (24,50) (25,50) (26,50)
  (27,50) (28,50) (29,50) (30,50) (31,50)
  (32,50)
};

\addplot[
  red,
  thick,
  dashed
] coordinates {(0,46.75) (32,46.75)}
node[pos=0.02, xshift=0ex, yshift=0.5ex, anchor=south west, text=red] {Worst-case threshold $t_{RC}$};

\end{axis}
\end{tikzpicture}
\caption{Maximum serviceable rate (inverse of $\overline{\delta_{p}}$) for batches of PCIe Read requests to $R$ unique rows in one bank.}
\label{fig:lookahead}
\end{figure}
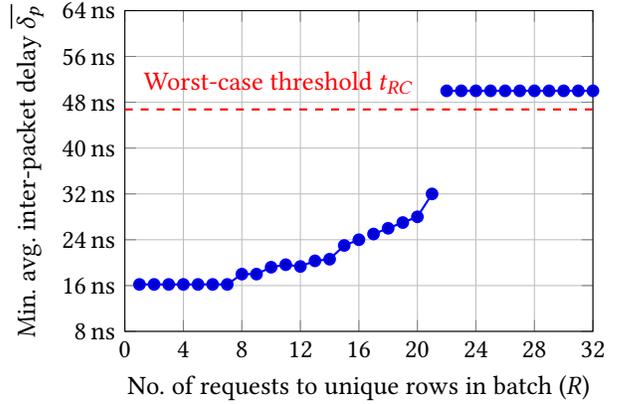

We follow on from our experiment that found the maximum stable transmission rate, where we repeatedly transmit one \pcie{Read} TLP to the same address with $\overline{\delta_{p}} = 16.2\,\text{ns}$. 
We alter our request pattern to instead send batches of \pcie{Read} TLPs to progressively larger sets of addresses ($R$) that access unique rows within the same bank.
\cref{fig:lookahead} shows the minimum serviceable inter-packet delay $\overline{\delta_{p}} = 1/\mu_{\text{out}}$ (corresponding to the maximum service rate $\mu_{\text{out}}$), for varying $R$.

In the worst-case scenario where there is no performance benefit from reordering, for large $R$ we expect each request to be served by a full \texttt{PRE} $\rightarrow$ \texttt{ACT} $\rightarrow$ \texttt{RD} $\rightarrow$ \texttt{PRE} sequence. 
Such a cycle is constrained by $t_{RC}$, which is the minimum interval between \texttt{ACT}s in a bank~\cite{jedec_ddr4_2020}.
For our target DIMM, this value is 46.75\,ns~\cite{M378Addr4}, which explains why any $R$ greater than those shown in the figure is also serviceable with $\overline{\delta_{p}}\approx50\,\text{ns}$.
We mark the $t_{RC}$ threshold on \cref{fig:lookahead}. 

On \cref{fig:lookahead} we immediately identify the abrupt jump from $R \geq 22$ where performance degrades to the worst-case with $\overline{\delta_{p}} > t_\text{RC}$. 
From this, we infer that the look-ahead window spans up to 21 entries.
For $R \leq 21$ we see a linear drop until $R=8$, and we note that performance values for $R<8$ are limited by our ZDMA PCIe connection bottleneck, as they are at the maximum stable transmission rate we previously found. 
We expect this drop to continue below $R<8$. 

Although observing the order of \pcie{Completion}s received in response to \pcie{Read} requests is not an entirely reliable indicator of memory controller reordering, our observations with this method corroborate our findings to this point. 
Moreover, using similar observations of \pcie{Completion}s for batch sizes susceptible to reordering ($R \leq 21$), we find that if we increase the inter-packet transmission interval $\overline{\delta_{p}}$, we observe a sharp transition when it goes beyond the per-bank activation cycle time $t_{\text{RC}}$ where no reordering occurs for all $R$. 

\begin{center}
    \fcolorbox{black}{gray!10}{\parbox{0.97\linewidth}{
    \textbf{Takeaway 2:}
    Sending successive PCIe requests with time intervals between each that are larger than the per-bank activation cycle time avoids reordering of memory requests.
    }}
\end{center}

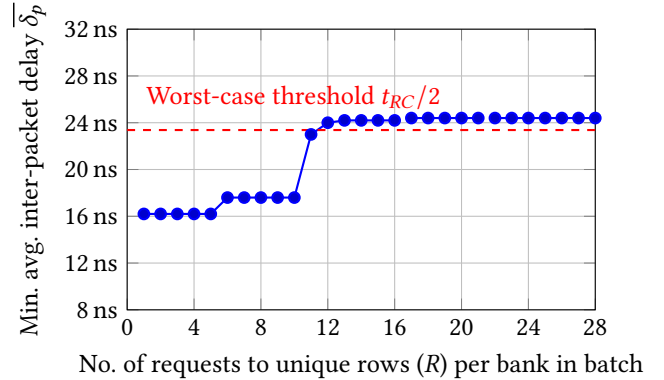
\begin{figure}[h]
\centering
\begin{tikzpicture}
\begin{axis}[
  scale only axis,
  width=0.35\textwidth,
  height=0.21\textwidth, 
  xmin=0, xmax=28,
  ymin=8, ymax=32,
  xlabel={No. of requests to unique rows ($R$) per bank in batch},
  ylabel={Min. avg. inter-packet delay $\overline{\delta_{p}}$},
  ytick distance=4,
  xtick distance=4,
  ymajorgrids,
  xmajorgrids,
  yticklabel=\pgfmathprintnumber{\tick}\,ns
]
\addplot+[
  mark=*,
  thick
] coordinates {
  
  (1,16.2) (2,16.2) (3,16.2) (4,16.2) (5,16.2)
  (6,17.6) (7,17.6) (8,17.6) (9,17.6) (10,17.6)
  (11,23) (12,24) (13,24.2) (14,24.2) (15,24.2) (16,24.2)
  
  (17,24.4) (18,24.4) (19,24.4) (20,24.4) (21,24.4)
  
  (22,24.4) (23,24.4) (24,24.4) (25,24.4) (26,24.4) (27,24.4) (28,24.4)  
};

\addplot[
  red,
  thick,
  dashed
] coordinates {(0,23.375) (28,23.375)}
node[pos=0.02, yshift=1ex, anchor=south west, text=red] {Worst-case threshold $t_{RC}/2$};

\end{axis}
\end{tikzpicture}
\caption{Maximum serviceable rate (inverse of $\overline{\delta_{p}}$) for batches of PCIe Read requests to $R$ unique rows to two banks.}
\label{fig:lookaheadmulti}
\end{figure}

\parhead{Multi-Bank Look-Ahead.}
We repeat the same experiment, this time with batches that access $R$ unique rows in each of $B=2$ different banks.
In this situation, the controller can leverage bank-level parallelism, which for $B=2$ we expect to roughly halve the worst-case performance scenario to take $\sim$23.375\,ns $(=t_{RC}/B)$.
The results, in \cref{fig:lookaheadmulti}, show  a similar jump in serviceable interval for total accesses $B \times R = 22$ (where $R = 11$). 
The slight drop for $R = 11$, compared to the flat line above the worst case performance threshold for $ R > 11$ can be explained by one of the bank's access patterns occasionally benefiting from reordering, corresponding to the $B \times R = 21$ case in \cref{fig:lookahead}.
Overall, this experiment confirms that the look-ahead buffer is not per-bank, but rather acts across the entirety of RPQ. 
We confirm this by repeating the experiment across banks in the same bank groups, and in different bank groups. 

We similarly confirm these findings with observations of the returning \pcie{Completion} flow. 
Curiously, however, we find that if the two banks are in different bank \emph{groups} there appears to be reordering evident in the order of returning \pcie{Completion}s for $R$ up to 25.
Nevertheless, this reordering does not correspond to an increase in performance, and in this case, we see the same results as in \cref{fig:lookaheadmulti}. 
We additionally confirm the findings summarised in Takeaway 2. 

\begin{center}
    \fcolorbox{black}{gray!10}{\parbox{0.97\linewidth}{
    \textbf{Takeaway 3:}
    The look-ahead window acts across all banks and can perform reordering on up to 21 entries.
    }}
\end{center}

\subsection{Bitflips in DDR4 via PCIe}\label{section:DDR4flips}
By using the ZDMA device with our modifications described in the previous sections, along with hammering transmission strategies that we found could avoid reordering, we are able to induce bitflips in targets with DDR4 system DRAM.
Specifically, we achieve this using the (multi-sided) Multibank hammering access patterns~\cite{kang2024sledgehammer} with a large number of aggressors that we profiled from software.

To recap, Multibank (from Sledgehammer~\cite{kang2024sledgehammer}) is primarily based on TRResspass~\cite{frigo2020trrespass}, which overcomes TRR countermeasures found in most DDR4 by performing multi-sided hammering that overwhelms the samplers keeping track of frequently accessed rows.
Multibank further improves on this by exploiting bank-level parallelism, which enables hammering across multiple banks $B$ at a rate constrained by $\sim$$B/t_{\text{RC}}$ instead of $\sim$$ 1/t_{\text{RC}}$. 
This has two overall beneficial effects: parallelising bitflip location search essentially ``for free''; and for $R<W$, filling the look-ahead window with other entries that minimise reordering, amounting to more effective per-bank aggressor sets. 
In the case of our target system, our experiments in \cref{subsection:reorderexps} show that the latter effect benefits any patterns with $R<22$, provided this is in combination with $B>1$. 
Our modified ZDMA hardware is well suited to take advantage of these effects for up to $B\leq3$, as from what we found, our ZDMA device's maximum transmission rate \hbox{($1/\overline{\delta_{p}} = 1/(16.2\,\text{ns})$)} which is close to the 3-bank hammering constrained rate of $B/t_{\text{RC}}=3/(46.75\,\text{ns})$.

\parhead{Optimal Hammering Timings For PCIe.}
We update our hammering-over-PCIe approach targeting DDR4 from \cref{section:DDR4attempt} to investigate various combinations of timing parameters, guided by our findings on how we can avoid reordering from the previous section.
Namely, in accordance with Takeaway 2, we transmit our batch of $B \times R$ aggressor PCIe \pcie{Read} TLP requests such that the average inter-packet delay results in a request rate that is slightly above the serviceable rate: $\overline{\delta_{p}} > B/t_{\text{RC}}$.

We use the same aggressor sets profiled with Multibank from software. 
Of the evaluated combinations of parameters, we measure the ``goodness'' of each one by the number of profiled bitflips observed which we can reproduce via PCIe. 

Initially, when we attempt to hammer with $\overline{\delta_{p}}$ just over $B/t_{\text{RC}}$, and no inter-batch delay, i.e., $\delta_{p} = \overline{\delta_{p}}$ we cannot observe any bitflips. 
However, once we introduce some inter-batch delay $\delta_{b}>0$ we finally observe bitflips in victim rows.
Specifically, as we increase this value we observe a spike in the number of bitflips induced right at the transition of $\delta_{b} > 2 \times t_{\text{RC}}$ (similarly $T_b \geq (B\times R+1)\times t_{\text{RC}}$). 
We suspect the extra delay of one $t_{RC}$ period (responsible for the $+1$ term) serves as a buffer drain relief and/or noise resilience period. 
This threshold is crucial for attack performance and appears to be constant as the same holds for any overall size of aggressor set $(B \times R)$ we use, and similarly for any bank cardinality $B$ of the set. 

Moreover, we find that if we keep $\delta_{b}$ above the mentioned threshold and reduce $\delta_{p}$ such that we still satisfy our reordering prevention condition: $\overline{\delta_{p}} > B/t_{\text{RC}}$ (which is possible due to the added $\delta_{b}$), we further increase the number of bitflips induced. 
The number of bitflips rises as ${\delta_{p}}$ shrinks until it passes below: ${\delta_{p}}<t_{\text{RC}}/B$.
By reducing the inter-packet delay just below this point, we ensure that the requests arrive at the memory controller in a burst such that we maximise \texttt{ACT} rate.
With further reduction below this point, we do not find there to be any attack performance improvement. 

An inter-packet delay $\delta_p$ of just below $t_{\text{RC}}/B$ corresponds to a request rate that is slightly faster than the maximum possible service rate that has an \texttt{ACT} for each request, this ensures that requests reach the memory controller in a burst. 

To summarise, we maximise the number of bitflips induced from hammering over PCIe when the timing of our aggressor transmissions meets the following three conditions:
\begin{enumerate}
    \item $\overline{\delta_{p}} > B/t_{\text{RC}}$:  Reordering prevention
    \item $\delta_{b} > 2 \times t_{\text{RC}}$: Noise robustness
    \item ${\delta_{p}}<t_{\text{RC}}/B$: Maximising \texttt{ACT} burst rate within a batch
\end{enumerate}

The best combinations we find can generally appear to reproduce all profiled bitflips.

\subsection{Bitflips in DDR4 via Thunderbolt}

We successfully reproduce a majority of the same bitflips in DDR4 targets when tunnelling our PCIe connection through Thunderbolt. 
In our setting, we use a Startech Thunderbolt\,3 PCIe Expansion Chassis (TB31PCIEX16) which has a x16 lane PCIe Gen 3 (8\,Gbps lanes) slot~\cite{startech_tb31pciex16}. 
However, due to Thunderbolt only using x4 lanes, our ZDMA connected through this chassis maximises lane utilisation. 

PCIe tunnelling through Thunderbolt appears to be an optional feature for systems to support. 
The DDR4 system we target, i.e., Desktop3 (~\cref{tab:platforms}), supports Thunderbolt\,3. 

Overall, we find that the attack tunnelled through Thunderbolt can reproduce roughly 80\% of the bitflips found with PCIe hammering.
Though there is some degradation in attack performance, our findings suggest that the intermediate chassis used for tunnelling does not impose an attack-inhibiting influence on PCIe request stream sent out for hammering.

\section{Evaluation}

\subsection{Platforms}
\begin{table*}[htbp]
\centering
\caption{Specifications of DDR3 and DDR4 Platforms}
\label{platforms}
\begin{tabular}{lllll}
\hlineB{2}
 \textbf{Platform} & \textbf{CPU Model} & \textbf{Motherboard} & \textbf{Memory Size/Model} & \textbf{TB} \\
\hlineB{2}
 Desktop1  & Intel Core i7-3615QM  & Apple Inc. Mac-F65AE981FFA204ED & 4\,GB DDR3 / M471B5273DH0-CH9  & Yes \\

Desktop2  & Intel Core i7-4790 & LENOVO SHARKBAY SDK0E50510 & 4\,GB DDR3 / M378B5273DH0-CH9 & No \\
 Desktop3 & Intel Core i7-7700 &  Gigabyte Z170X-Gaming 7 & 8\,GB DDR4 / M378A1K43BB1-CPB & Yes \\
\hline
\end{tabular}
\label{tab:platforms}
\end{table*}

\cref{tab:platforms} summarises all DDR3 and DDR4 platforms we experiment on.
Among them, Desktop1 and Desktop3 support Thunderbolt, which is a common feature on modern laptops.
For Thunderbolt chassis, our DDR3 setup uses the Sonnet Echo Express SE II, while the DDR4 setup uses the Startech Thunderbolt 3 PCIe Expansion Chassis, both of which are also discussed in \cref{sec:simplehammer} and \cref{sec:mcre}.
Although we explored both DDR3 and DDR4 systems, our evaluation hereon focuses on DDR4 systems, which are more widespread and for which bitflips are harder to trigger.

\subsection{Case Study with IOMMU enabled}

Our investigation until this point primarily focuses on the practicality of generic PCIe and Thunderbolt peripherals serving as an attack vector for Rowhammer. 
Having demonstrated the feasibility of this approach, we now extend our analysis to investigate whether bitflips induced from PCIe and Thunderbolt accesses are still possible on systems with memory virtualisation and the IOMMU enabled. 
Until now, our testing involved direct interaction with physical memory addresses, with memory virtualisation disabled in our targets.
This adds two more potential stages of processing between our source of hammering PCIe requests and the targeted DRAM. 
Namely, the host must translate the virtual addresses of incoming requests to the corresponding physical memory addresses, then its IOMMU must perform necessary bounds checks.
We find that even under these circumstances, we can cause bitflips, which indicates that neither of these extra stages inhibit Rowhammer.

When memory virtualisation and the IOMMU are enabled in a host system, any device connecting over PCIe must be explicitly allocated a memory region on the host which it can read and write to. 
Soon after device connection or system startup, once the device driver is loaded and has been allocated a region for device DMA, the driver passes the region base address and size to the device. 
The host communicates this by writing to one of the device's Base Address Registers (BAR). 
As we mention earlier, pcileech does not support operation on systems with virtualisation enabled.
The reason for this is the codebase does not support reading out the data that hosts write to the device BARs. 
We further modify our custom ZDMA hardware to enable BAR readout using the pcileech control software. 

\parhead{Setup.}
To demonstrate \myname{} works under IOMMU protection, we enable virtualisation and the IOMMU by enabling \texttt{VT-d} in BIOS and updating the GRUB configuration with the line \texttt{intel\_iommu=on}.
For simplicity in testing, we create a custom kernel-space driver on the host to first enable our ZDMA device connection through Thunderbolt.
Using \texttt{dma\_declare\_coherent\_memory} we then allocate the DMA regions that correspond to aggressor rows through which we previously found hammering could cause bitflips. 
Next, we allocate the DMA region to the device with \texttt{dma\_alloc\_coherent}, which handles writing the region address offset and size to the device BAR. 
Finally, our driver enables the DMA area.
We use code running on the host to observe whether the victim rows, which are outside of the device's allocated region, experience any bitflips.

\parhead{Bitflip Reproduction Using PCIe and Thunderbolt.}
After driver initialisation, we attempt to hammer with our customised ZDMA by sending \texttt{Read} TLP packets to the host machine.
We similarly find approximately 80\% of bitflips found to be reproducible with this approach.

\parhead{Thunderhammer and Software-based Mitigation.}
Several Rowhammer mitigation schemes are geared to thwart hammering from software by monitoring potentially malicious memory access patterns~\cite{zhang2022softtrr}, or attack detection by monitoring performance counters~\cite{seaborn2015exploiting, aweke2016anvil}. 
In such cases where code running on the host is unable to hammer, a malicious driver in collusion with Thunderhammer may be capable of mounting attacks.

\section{Limitations, Future Work and Countermeasures.}

\subsection{Using Newer PCIe Generations.}
We expect that adopting our attack approach to newer PCIe generation devices will increase the attack effectiveness.
In our attacks we use our modified ZDMA device that connects over x4 lanes of PCIe Gen 2, each of which operates at 5\,Gbps.
As we discuss in \cref{section:DDR4flips}, this device can transmit a \pcie{Read} TLP every 16.2\,ns on average.
This is well below $t_{RC}$, which allows us to maximise the effective approach for single-bank hammering. 
This is close to the 3-bank hammering constrained rate $B/t_{\text{RC}}=3/(46.75\,\text{ns})$, so the attack can leverage Multibank hammering for bank cardinality $B$ up to 3. 
However, 
Thunderbolt versions 3 and 4 utilise x4 PCIe Gen 3 lanes for tunnelling, which each have a bitrate of 8\,Gbps.
We expect using a PCIe Gen 3 device to increase the effectiveness of Multibank hammering for higher bank cardinality. 
Similarly, we expect further benefits for the direct PCIe slot attack approach, similar function can be built-in to either a device from a much more recent PCIe generation (4 or 5) or devices with more lanes, e.g. x16, can be used.

Crucially, we expect these improvements to carry into newer generations of all the technologies such as PCIe Generations 6.0 and onward, whose transmission rates are over an order of magnitude greater than those we experiment with. 
This may have implications on the security of more robust, recent memory systems such as DDR5.
We leave the development of hammering devices using newer PCIe generations to future work.

\subsection{Attacks on Server-Grade CPUs}

Our study focuses on \myname{} attacks targeting client-grade CPUs, i.e., Intel Core machines. 
We expect that these attacks may not directly transfer to server-grade CPUs, such as Intel Xeon machines. 
These systems serve PCIe-initiated memory accesses from Last-Level Cache (LLC) using mechanisms like Data Direct I/O (DDIO)~\cite{intelDDIO}.

Unlike software-initiated memory accesses on the CPU side, there is no mechanism for flushing the cache, such as with a \texttt{clflush} instruction, from a PCIe connection. 
However, we expect that by constructing and accessing eviction sets, our \myname{} approach may also be adaptable to attacking these platforms.
We leave the investigation of attacks on server-grade CPUs to future work. 

\subsection{Mitigations.}

\parhead{General Rowhammer Mitigations.} Strategies effective against conventional Rowhammer attacks may mitigate \myname{} to some extent.
Although ECC cannot fully mitigate Rowhammer, it could significantly reduce bitflips. 
Based on ECC, Copy-on-Flip\cite{di2023copy} proposes a software-based solution that migrates the victim data to new memory pages.

Moreover, some DRAM modules are more susceptible to Rowhammer-induced bitflips than others, as disclosed in previous studies~\cite{frigo2020trrespass}. 
Thus, for systems running critical applications, defenders can avoid DRAM chips known to be vulnerable. 
Besides, they can actively identify vulnerable chips by searching for bitflips using tools such as Blacksmith~\cite{jattke2022blacksmith},  BitMine~\cite{zhang2021bitmine}, and Rowhammer-test~\cite{google_rowhammer_test}.

Another straightforward Rowhammer mitigation is to double the DRAM refresh frequency, which reduces the likelihood of disturbance errors. 
However, this approach increases power consumption and introduces performance overhead, making it a trade-off. 
This solution has already been adopted in DDR5, where the refresh interval was reduced from 64 ms to 32 ms~\cite{jattke2024zenhammer}.

\parhead{Limiting TLP Request Rates.} 
To mitigate \myname{}, we can also introduce mechanisms to limit the TLP request rate of PCIe peripherals, especially suspicious ones. 
Such a mechanism could be implemented in the PCIe Root Complex or PCIe switches. 
Note that even if the request rate is limited, the bandwidth can remain unchanged, as a peripheral may send fewer TLPs with larger payloads.

\section{Conclusions}
\label{section:conclusion}

In this work, we present \myname{}, the first general PCIe- and Thunderbolt-based Rowhammer attack capable of inducing DRAM bitflips in modern DDR4 targets. 
By assessing how memory access patterns are influenced by memory controller buffering, we show that carefully timing-tuned PCIe transactions can reliably launch Rowhammer attacks on DDR3 and DDR4 systems. 
This previously unexplored peripheral-based vector broadens the Rowhammer threat surface and underscores the need for Rowhammer defences against a wide range of attack vectors. 
  
\section*{Acknowledgements}
This work was supported by 
an ARC Discovery Project number DP210102670; 
the Deutsche Forschungsgemeinschaft (DFG, German Research Foundation) under Germany's Excellence Strategy - EXC 2092 CASA - 390781972; 
and the National Science Foundation (Grant No. CNS-2145744).

%\bibliographystyle{ACM-Reference-Format}
%\bibliography{references.bib}
%%% -*-BibTeX-*-
%%% Do NOT edit. File created by BibTeX with style
%%% ACM-Reference-Format-Journals [18-Jan-2012].

\end{document}